\documentclass[superscriptaddress,twocolumn,prd,aps,showpacs,secnumarabic]{revtex4}
\usepackage{graphicx}
\begin{document}

\title{Higgs boson decay into $Z$ bosons and a photon}
\author{Ali Abbasabadi}
\affiliation{Department of Physical Sciences, Ferris State University, Big Rapids, Michigan 49307, USA}
\author{Wayne W. Repko}
\affiliation{Department of Physics and Astronomy, Michigan State University, East Lansing, Michigan 48824, USA}

\date{\today}
\begin{abstract}
\vspace*{0.5cm} \hspace*{0.1cm} We have performed a one-loop calculation of the width of the rare decay $H\to Z Z \gamma$ in the standard model for Higgs boson masses $190\, {\rm GeV} \leq m_H \leq 250\, {\rm GeV}$. We find that the most dominant helicity combinations for the $Z$ bosons and the photon is when one of the $Z$ bosons is longitudinally polarized and the other $Z$ boson and the photon have the same helicity. A comparison of the decay width $\Gamma(H\to Z Z \gamma)$ to those of $H\to\gamma \gamma$ and $H\to\gamma Z$ shows that the ratios of the decay widths are $\Gamma(H\to Z Z \gamma) / \Gamma(H\to\gamma\gamma) \sim \Gamma(H\to Z Z \gamma) / \Gamma(H\to \gamma Z) \lesssim 10^{-7}$.
\end{abstract}
\pacs{13.15.+g, 14.60.Lm, 14.70.Bh, 95.30.Cq} \maketitle

\vskip1pc

\section{Introduction}
\label{sec:1} In the standard model, the lowest order contribution to the decay $H\to Z Z \gamma$ takes place at the one-loop level. The Feynman diagrams for this process are similar to those of the rare decay process $H\to\gamma\gamma Z$\cite{ar}. The decays $H\to\gamma\gamma$ and $H\to\gamma Z$ which also occur at the one-loop level, and $H\to Z Z$ which occurs at the tree level, dominate the decay process $H\to Z Z \gamma$ by several orders of magnitude\,\cite{ghkd}. However, by imposing kinematic cuts on the $Z$ bosons and the photon in the decay $H\to Z Z \gamma$, we may exclude contributions of the back-to-back $Z$ bosons, and distinguish the $Z$ bosons of the $H\to Z Z \gamma$ decay from those of the $H\to Z Z$ decay. Although very rare, the decay $H\to Z Z \gamma$ is sensitive to the top quark couplings, and therefore, a signal in this channel is an evidence for a deviation from the standard model top quark couplings.

In the next section, we outine our calculations of the decay width, the $Z$ boson invariant mass decay distribution, and the photon's energy decay distribution. This is followed by a summary and conclusions.

\section{Decay Width Calculation}
\label{sec:2}
In order to decrease the number of the Feynman diagrams needed to calculate the amplitudes for the process $H\to Z Z \gamma$, we use the generalized non-linear gauge fixing condition introduced in the Ref. \cite{bc}. The parameters contained in the gauge fixing terms do not affect observables and may be chosen for convenience. We choose the values of these parameters in such a way as to eliminate certain vertices and thereby minimize the number of diagrams to be calculated. To further simplify the structure of the resulting tensor integrals and to increase their numerical stability, we use the 't Hooft-Feynman gauge in which the propagators for the gauge bosons have the form $-ig_{\mu\nu}/(k^2 - m^2)$, where $k$ and $m$ are the momentum and the mass of a gauge boson, respectively.

\begin{figure}[h]
\centering\includegraphics[width=2.5in]{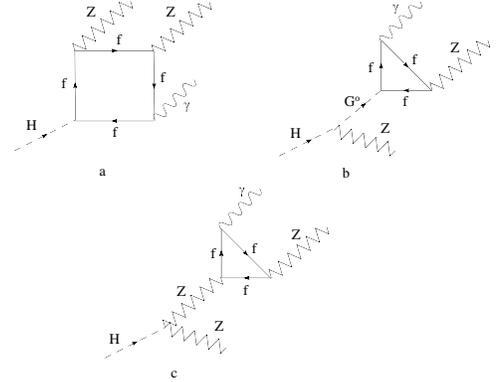}
%\vspace{.10in}
\caption{\footnotesize Some representative Feynman diagrams for the decay process $H\to Z Z \gamma$ are shown. For the charged fermion $f$ in the Figs.\,\ref{feyndiag}(a) and \ref{feyndiag}(b), we include only the top quark. In the Fig.\,\ref{feyndiag}(c), all charged fermions $f$ of the third generation are included.} \label{feyndiag}
\end{figure}

The resulting Feynman diagrams are similar to those of the rare decay process $H\to\gamma\gamma Z$\,\cite{ar} and its crossed channel scattering process $\gamma\gamma\to Z H$\,\cite{gpr}. Some representative diagrams are drawn \cite{jaxo} in the Fig.\,\ref{feyndiag}.
In Ref. \cite{gpr} it was noted that the charge conjugation properties of the gauge boson couplings exclude any contribution from $W$ bosons in the loops and only charged fermions in the loops contribute. In the Fig.\,\ref{feyndiag}(a), the coupling of the Higgs boson to the fermion is proportional to the fermion mass and we include only the top quark contribution. The same is true in the Fig.\,\ref{feyndiag}(b) for the coupling of the neutral Goldstone boson $G^0$ and the fermion, and again we include only the top quark contribution.

The contribution of the diagram of the Fig.\,\ref{feyndiag}(c), apart from the dependence on the mass of the fermion in the loop, is proportional to $N^f_{c}\,Q_f \, g^f_{A}\, g^f_{V}$. Here, $N^f_{c}$ is the number of fermion colors ($N^f_{c} = 1$ for a lepton, $N^f_{c} = 3$ for a quark), $Q_f$ is the fermion electric charge in units of the proton charge, $g^f_{A} = T^f_{3}$ and $g^f_{V} = T^f_{3} - 2 Q_f\, \sin^2 \theta_W$ are the axial-vector and the vector coupling constants, respectively.  $T^f_3$ is the third component of the weak isospin, and $\theta_W$ is the weak mixing angle. As far as its dependence on the fermion mass is concerned, the contribution of the diagram of the Fig.\,\ref{feyndiag}(c) consists of two parts. One depends on the mass of the charged fermion in the loop (it vanishes for a massless fermion), and the other is independent of the fermion mass. The latter gives an anomalous contribution \cite{a}. However, it is clear that the inclusion of all charged fermions of a given generation will cancel this anomalous contribution since $\sum_f N^f_{c}\, Q_f\, g^f_{A}\, g^f_{V} = 0  \,$\cite{note1}. Furthermore, if all of the members of a particular generation had the same mass, the total contribution of that generation would vanish. We included only the charged fermions of the third generation in the evaluation of the diagram of the Fig.\,\ref{feyndiag}(c). (The other two generations give negligible contributions.)

The process $H\to Z Z \gamma$ will be dominated by the decays $H\to\gamma\gamma$, $H\to\gamma Z$, and especially $H\to Z Z$ that occurs at tree level. In order to facilitate the discrimination of $H\to Z Z \gamma$ from these dominant decay modes and account for some of the possible experimental limitations, we imposed cuts on the following $Z$ boson and photon kinematic variables: $|\vec{p}_Z|$, $|\vec{p}_{Z'}|$, $|\vec{p}_\gamma|$, $m_{Z Z'}^2$, $m_{\gamma  Z}^2$, $m_{\gamma Z'}^2$, $\theta_{Z Z'}$, $\theta_{\gamma Z}$, and $\theta_{\gamma Z'}$. Here, $\vec{p}_Z$, $\vec{p}_{Z'}$, and $\vec{p}_{\gamma}$ are the 3-momenta of the $Z$ bosons and photon, respectively, in the center of mass of the Higgs boson, and $\theta_{Z Z'}$, $\theta_{\gamma Z}$, and $\theta_{\gamma Z'}$ are the various angles between the 3-momenta, $\vec{p}_Z$, $\vec{p}_{Z'}$, and $\vec{p}_{\gamma}$. The invariant mass variables are $m_{Z Z'}^2 = (p_{Z}+p_{Z'})^2$, $m_{\gamma Z}^2 = (p_{\gamma}+p_{Z})^2$, and $m_{\gamma Z'}^2 = (p_{\gamma}+p_{Z'})^2$, where $p_{Z}$, $p_{Z'}$, and $p_{\gamma}$ are the 4-momenta of the $Z$ bosons and the photon.

We choose the following set of cuts for the numerical calculations of the decay width:
\begin{eqnarray}
|\vec{p}_{Z}|_{\rm cut}= |\vec{p}_{Z'}|_{\rm cut}=
|\vec{p}_{\gamma}|_{\rm cut} &\equiv& |\vec{p}\,|_{\rm cut} \,,\label{pcut}\\
(m_{Z Z'})_{\rm cut}= (m_{\gamma Z})_{\rm cut}=
(m_{\gamma Z'})_{\rm cut} &\equiv&  m_{\rm cut}\,,\label{mcut}\\
(\theta_{Z Z'})_{\rm cut}= (\theta_{\gamma Z})_{\rm cut}= (\theta_{\gamma Z'})_{\rm cut} &\equiv& \theta_{\rm cut} \,.\label{tcut}
\end{eqnarray}
These kinematic cuts facilitate the experimental tagging of the $Z$ bosons and photon. They provide minimum opening angles between the $Z$ bosons and the photon, exclude contributions of the back-to-back $Z$ bosons and photons, exclude soft photons, and also improve the numerical stability of the calculations. The cuts help discriminate the non-back-to-back $Z$ bosons and the photon of the decay $H\to Z Z \gamma$ from the back-to-back $Z$ bosons and photons in the decays $H\to Z Z$, $H\to \gamma Z$, and $H\to \gamma\gamma$. In principle, all the $Z$ bosons and photons of the decays $H\to Z Z \gamma$, $H\to Z Z$, $H\to \gamma Z$, and $H\to \gamma\gamma$ can be identified. However, resolving ambiguities in the reconstruction of the $Z$ bosons can be problematic, especially when decays into neutrino-antineutrino pairs are involved.

For the calculation of the decay width, and its distributions with respect to the invariant mass and energy, we have used a {\it semiautomatic} \cite{hs} method. In this method, we used the FeynArts package \cite{h} to generate the helicity amplitudes in terms of the tensor loop integrals and then, using the Passarino-Veltman method \cite{pv}, we expressed these tensor integrals in terms of scalar integrals and evaluated them numerically \cite{note2}.

\begin{figure}[h]
\centering\includegraphics[width=2.5in]{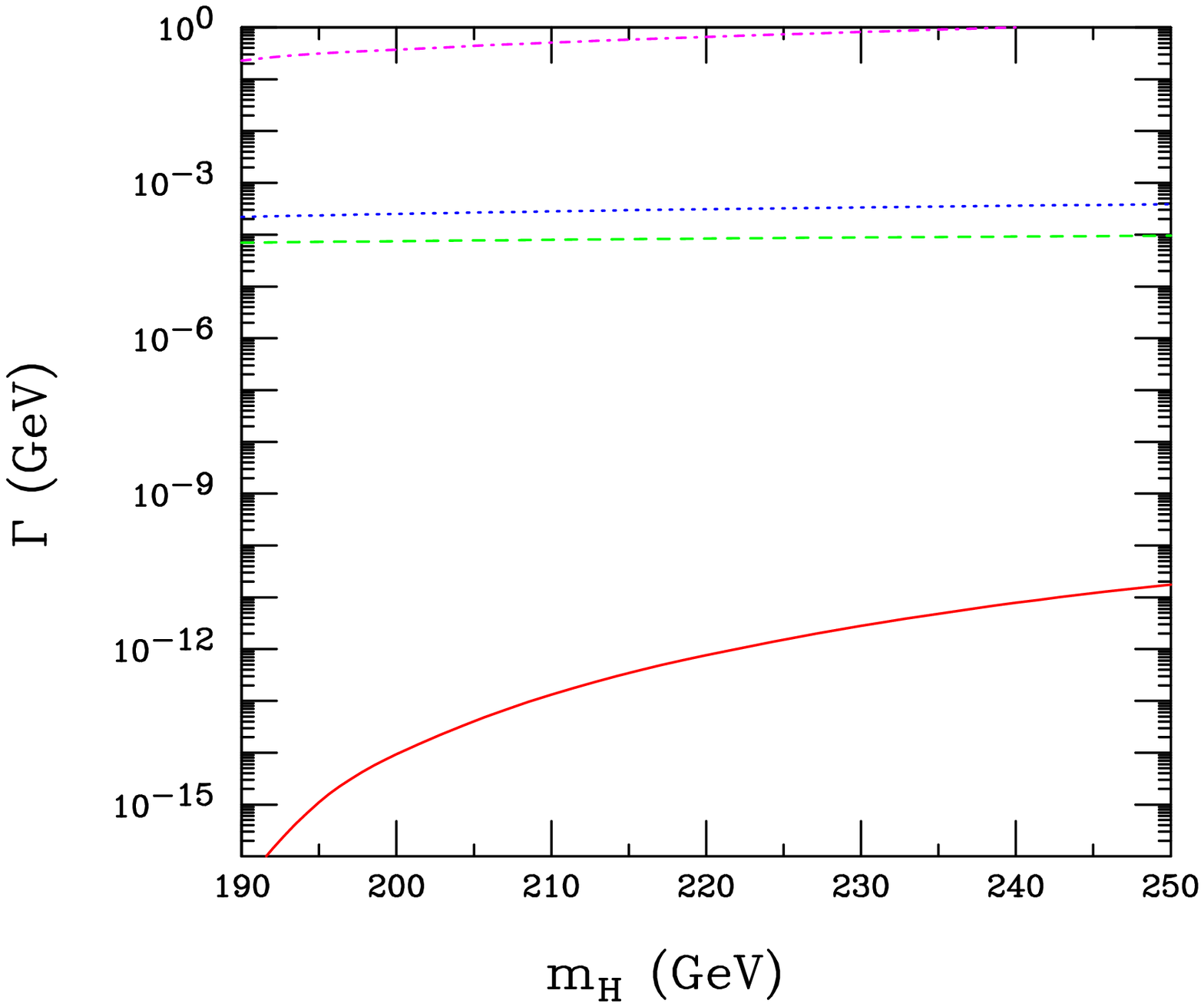} \caption{\footnotesize (Color online) The decay widths as function of  $m_H$ for several decay modes of the Higgs boson are shown. The solid line is $\Gamma(H\to Z Z \gamma)$, the dashed line is $\Gamma(H\to\gamma\gamma)$, the dotted line is $\Gamma(H\to\gamma Z)$, and the dotdashed line is $\Gamma(H\to Z Z)$. The cuts imposed on $\Gamma(H\to Z Z \gamma)$ are $|\vec{p}\,|_{\rm cut}= 5\, \rm{GeV}$, $m_{\rm cut}= 10\, \rm{GeV}$, and $\theta_{\rm cut} = \pi/24$.} \label{decaywidth}
\end{figure}

The result of the calculation for the decay width $\Gamma(H\to Z Z \gamma)$ as function of the Higgs boson mass $m_H$ is shown in the Fig.\,\ref{decaywidth}.
For comparison, in this figure we also included the decay widths $\Gamma(H\to Z Z)$, $\Gamma(H\to \gamma\gamma)$, and $\Gamma(H\rightarrow \gamma Z)$ \cite{vvss,hdecay}. It is clear from this figure that the decay width $\Gamma(H\to Z Z \gamma)$ is several orders of magnitude smaller than those of $H\to Z Z$, $H\to \gamma\gamma$, and $H\to \gamma Z$. For Higgs boson masses $190\, {\rm GeV} \leq m_H \leq 250\, {\rm GeV}$, the ratios of the decay widths are $\Gamma(H\to Z Z \gamma) / \Gamma(H\to \gamma\gamma) \sim \Gamma(H\to Z Z \gamma) / \Gamma(H\to\gamma Z) \lesssim 10^{-7}$ and $\Gamma(H\to Z Z \gamma) / \Gamma(H\to Z Z) \lesssim 10^{-11}$. To identify the origins of the smallness of these ratios, we note that, in addition to the cuts that we imposed on the decay products of $H\to Z Z \gamma$, which decrease the value of $\Gamma(H\to Z Z \gamma)$, there is also the suppression from three-body phase space, and from the higher order in the coupling constant $\alpha$. These, however, do not completely account for the suppression. There are other differences in the various decay amplitudes, which contribute to the small ratios. For instance, in the case of $H\to \gamma\gamma$, the decay amplitude receives contributions from charged fermion loops as well as a substantial contribution from $W$ boson loops, whereas in the decay $H\to Z Z \gamma$, there are no $W$ boson loop contributions and the inclusion of anomalous triangle diagram, Fig.\,\ref{feyndiag}(c), further suppresses the amplitude. As a result of these differences, the simple power counting method for estimating the size of the ratio of the decay widths $\Gamma(H\to Z Z \gamma) / \Gamma(H\to \gamma\gamma)$ is rather unreliable.

To investigate the dependence of the decay width $\Gamma(H\to Z Z \gamma)$ on the helicities of the produced $Z$ bosons and the photon, we can use Bose symmetry and the $CP$ invariance to obtain relations among the helicity amplitudes ${{\cal A}_{\lambda \lambda' \lambda_{\gamma}}}$. Here, $\lambda$ and $\lambda'$ are the helicities of the $Z$ bosons and $\lambda_{\gamma}$ is the helicity of the photon. As a consequence of these symmetries, the decay width $\Gamma_{\lambda \lambda' \lambda_\gamma}$ satisfies the following relations
\begin{eqnarray}
\Gamma_{\lambda \lambda' \lambda_\gamma}
= \Gamma_{\lambda' \lambda \lambda_\gamma} \,,\label{sym1}\\
\Gamma_{\lambda \lambda' \lambda_\gamma} = \Gamma_{-\lambda -\lambda' -\lambda_\gamma} \,.\label{sym2}\
\end{eqnarray}

\begin{figure}[h]
\centering\includegraphics[width=2.5in]{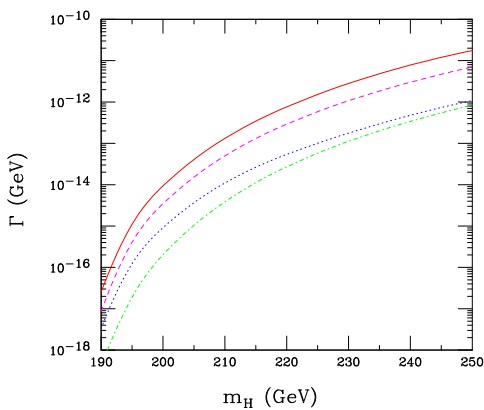} \caption{\footnotesize (Color online) The decay widths $\Gamma(H\to Z Z \gamma)$ as function of $m_H$ for different helicities $(\lambda \lambda' \lambda_\gamma)$ of the $Z$ bosons and the photon are shown. The solid line is for the unpolarized case, the dashed line is for $(0++)$, the dotted line is for $(0+-)$, and the dotdashed line is for $(00+)$.  The cuts imposed are the same as those for the total width $\Gamma(H\rightarrow ZZ\gamma)$ of the Fig.\,\ref{decaywidth}.} \label{decaywidthhel}
\end{figure}

In Fig.\,\ref{decaywidthhel}, we show the result of the calculation for the decay width $\Gamma(H\rightarrow Z Z\gamma)$ as function of the Higgs boson mass $m_H$, for different helicities of the $Z$ bosons and the photon.
As it is clear from this figure, the decay widths $\Gamma_{0++} = \Gamma_{0--}$, which correspond to the case when one of the $Z$ bosons is longitudinally polarized and the other $Z$ boson has the same helicity as that of the photon, are the most dominant. This dominance is stronger for the higher Higgs boson masses. The decay widths $\Gamma_{+++} = \Gamma_{---}$, $\Gamma_{++-} = \Gamma_{--+}$, and $\Gamma_{+-+} = \Gamma_{-+-} = \Gamma_{-++} = \Gamma_{+--}$, that are not shown in the Fig.\,\ref{decaywidthhel}, are negligible.  This pattern of the polarization states may be viewed as a signature for the decay products of the process $H\to Z Z \gamma$, since a longitudinal-transverse helicity combination cannot occur for the $Z$ pair from $H\to ZZ$.

\begin{figure}[h]
\centering\includegraphics[width=2.5in]{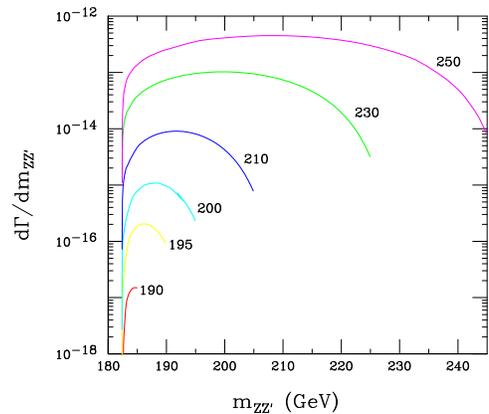} \caption{\footnotesize (Color online) The invariant mass distributions $d\Gamma(H\rightarrow Z Z \gamma)/dm_{ZZ'}$ as function of $m_{ZZ'}$, the invariant mass of the final $Z$ bosons, for Higgs masses of $m_H$ = 190, 195, 200, 210, 230, and 250 GeV are shown. The cuts imposed are the same as those for the total width $\Gamma(H\rightarrow ZZ\gamma)$ of the Fig.\,\ref{decaywidth}.} \label{massdist}
\end{figure}

\begin{figure}[h]
\centering\includegraphics[width=2.5in]{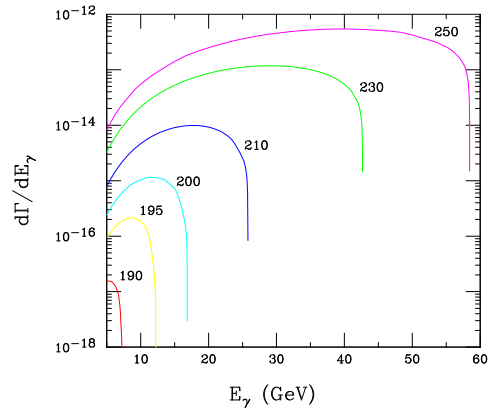} \caption{\footnotesize (Color online) The energy distributions $d\Gamma(H\rightarrow ZZ\gamma)/dE_{\gamma}$, as function of the photon energy $E_{\gamma}$, for the Higgs masses of Fig.\,\ref{massdist} are shown. The cuts imposed are the same as those for the total width  $\Gamma(H\rightarrow ZZ\gamma)$ of the Fig.\,\ref{decaywidth}.} \label{energydist}
\end{figure}

In Fig.\,\ref{massdist}, we show the invariant mass distribution $d\Gamma(H\to Z Z \gamma)/dm_{Z Z'}$ as function of the $Z$ boson pair invariant mass $m_{Z Z'}$, and in Fig.\,\ref{energydist}, we show the energy distribution $d\Gamma(H\to Z Z \gamma)/dE_{\gamma}$ as function of the photon energy $E_\gamma$. Notice that the study of the invariant mass and energy distributions for the Higgs boson masses of about $190\, {\rm GeV}$ is difficult. This is due to the imposed cuts that make the phase space quite small.

\section{Summary and Conclusions}
In the standard model, the three-body decay of the Higgs boson $H\to Z Z \gamma$ is highly suppressed. However, this decay mode has some interesting features that separate it from decay modes such as $H\to \gamma\gamma$ and $H\to \gamma Z$. One is the absence of $W$ boson contributions in any of the loops in the Feynman diagrams for the $H\to Z Z \gamma$. Its amplitudes are dominated by top quark loops and therefore sensitive to top-$Z$ couplings. Also, there is the presence of an anomalous vertex in the $s$-channel $Z$ exchange diagram of the Fig.\,\ref{feyndiag}(c), which might be studied were it not for the smallness of the decay width.

Our explicit calculations show that the most dominant helicity combinations for $H\to Z Z \gamma$ occur when one of the $Z$ bosons is longitudinally polarized and the other $Z$ boson and the photon have the same helicity. This is a result that was not apparent at the outset. With enough statistics, this feature might be used to discriminate the $Z$ pairs of $H\to ZZ\gamma$ from those of $H\to ZZ$, since the helicities of the $Z$ pair in the latter decay cannot be in the combination longitudinal-transverse.

In summary, we find that the decay width for the process $H\to Z Z \gamma$ is exceedingly small compared to those of  $H\to\gamma\gamma$ and $H\to\gamma Z$, and that the suppression is greater than the simple phase space and coupling constant accounting might suggest. The experimental measurement of the decay width for $H\to Z Z \gamma$ will be extremely difficult, especially the identification of the helicities of the $Z$ bosons and photon. Therefore, we expect not to detect a signal of this mode at the standard model level, and any detection is an indication of new physics beyond the standard model.

\begin{acknowledgements}
 One of us (A.A.) wishes to thank the Department of Physics and Astronomy at Michigan State University for its hospitality and computer resources. He would also like to thank Thomas Hahn for his advice and assistance in the implementation of the FeynArts package. This work was supported in part by the National Science Foundation under Grant No. PHY-02744789.
\end{acknowledgements}


\begin{thebibliography}{99}
\bibitem{note1} Notice that for the decay $H\to\gamma\gamma Z$ (calculated in Ref.\,\cite{ar}), the corresponding diagram in the Fig.\,\ref{feyndiag}(c) gives an anomalous contribution that is proportional to $N^f_{c}\,Q^2_f \, g^f_{A}$, and when all charged fermions of a given generation are included in the loop, this anomaly will also vanish, $\sum_f N^f_{c}\, Q^2_f\, g^f_{A} = 0$.
\bibitem{note2} We numerically checked the gauge invariance of the helicity amplitudes for the final state photon. The diagrams of the Figs.\,\ref{feyndiag}(a), \ref{feyndiag}(b), and \ref{feyndiag}(c) are separately gauge invariant. See Ref.\,\cite{ar} for a discussion and references regarding the calculation and test of the scalar integrals.
\end{thebibliography}
\end{document}